# Investigate the effect of anisotropic order parameter on the specific heat of anisotropic two-band superconductors

P.Udomsamuthirun[1,2], R.Peamsuwan[1] and C.Kumvongsa[3]

(1) Department of Physics, Faculty of Science, Srinakharinwirot University, Bangkok 10110,Thailand. E-mail: udomsamut55@yahoo.com
(2) Thailand Center of Excellence in Physics, Huay Kaew Road, Chiang Mai, Thailand, 50200
(3) Department of Basic Science, School of Science, The University of the Thai Chamber of Commerce, Dindaeng, Bangkok 10400,Thailand.

**Keywords:** the specific heat , two-band superconductors, anisotropic superconductors

**PACS** : 65.40.Ba, 74.20.Fg, 74.70 Ad.

**Abstract**

The effect of anisotropic order parameter to specific heat of anisotropic two-band superconductors in BCS weak-coupling limit are investigated. An analytical specific heat jump and the numerical specific heat are shown by using anisotropic order parameters, and the electron-phonon interaction and non-electron-phonon interaction. The two model of anisotropic order parameters are used for numerical calculation that we find little effect to the numerical results. The specific heat jump of $MgB_2$, $Lu_2Fe_3Si_5$ and $Nb_3Sn$ superconductors can fit well with the both of them. By comparing to experimental data overall range of temperature, the best fit is $Nb_3Sn$, $MgB_2$, and $Lu_2Fe_3Si_5$ superconductors, respectively.

## 1. Introduction

According to the discovery of the superconductor in $MgB_2$ [1], $Nb_3Sn$ [2-3] and $Lu_2Fe_3Si_5$ [4] have attracted the interesting in non-oxide superconductor and two-band superconductors. These superconductors are the good examples to study the two-band superconductors. The experimental measurements on specific heat of these superconductors are done by many researchers [5-7]. Recently, there are many researchers that proposed models in vicinity of two-band model to describe the specific heat of two-band superconductors. Palistrant and Ursu[8] described the specific heat jump of $MgB_2$ superconductor in the point $T = T_c$ by using the two-band model with reduced density charge carriers. Mishonov et al. [9] studied thermodynamics of $MgB_2$ superconductor by the weak-coupling two-band BCS model that they also showed the temperature dependent of specific heat. Guritanu et al [10] presented specific heat data for $Nb_3Sn$ superconductor and described by using the power expansion of the low-temperature lattice specific heat. Nagajima, Li and Tamegai[11] reported and described the low temperature specific heat of $Lu_2Fe_3Si_5$ superconductor by the phenomenological two-gap model. The two-band model was first introduced by Suhl[12] and Moskalenko[13]. And recently, Udomsamuthirun et al.[14] proposed the two-band model including electron-phonon interaction and an non-electron-phonon interaction and applied to calculate the isotope effect exponent of $MgB_2$ superconductor.

The propose of this paper is to apply the two-band model including electron-phonon interaction and an non-electron-phonon interaction [14] and anisotropic order parameter[15-17] to calculate the specific heat of two-band superconductors. Finally, we compare our calculations to the experimental data of $MgB_2$, $Lu_2Fe_3Si_5$ and $Nb_3Sn$ superconductors.

## 2. Model and calculation

The Hamiltonian of two-band system is taken in the form of linear combination of BCS Hamiltonian of first band, second band and interband so that after performing a BCS mean field analysis, we can obtain the gap equation as[14]

$$\Delta_{1k} = -\sum_{k'} V_{11kk'} \frac{\Delta_{1k'}}{2\sqrt{\varepsilon_{1k'}^2 + \Delta_{1k'}^2}} \tanh(\frac{\sqrt{\varepsilon_{1k'}^2 + \Delta_{1k'}^2}}{2T}) - \sum_{k'} V_{12kk'} \frac{\Delta_{2k'}}{2\sqrt{\varepsilon_{2k'}^2 + \Delta_{2k'}^2}} \tanh(\frac{\sqrt{\varepsilon_{2k'}^2 + \Delta_{2k'}^2}}{2T})$$

$$\Delta_{2k} = -\sum_{k'} V_{22kk'} \frac{\Delta_{2k'}}{2\sqrt{\varepsilon_{2k'}^2 + \Delta_{2k'}^2}} \tanh(\frac{\sqrt{\varepsilon_{2k'}^2 + \Delta_{2k'}^2}}{2T}) - \sum_{k'} V_{12kk'} \frac{\Delta_{1k'}}{2\sqrt{\varepsilon_{1k'}^2 + \Delta_{1k'}^2}} \tanh(\frac{\sqrt{\varepsilon_{1k'}^2 + \Delta_{1k'}^2}}{2T})$$

(1)

where $\Delta_{1k}$ and $\Delta_{2k}$ are the superconducting order parameters of $1^{st}$ band and $2^{nd}$ band that dependent on wave vector, $\vec{k}$ .

We make the assumption that each paring interaction potential consisted of 2 parts[14,18,19] : an attractive electron-phonon interaction $V_{ph}$ and an attractive non-electron-phonon interaction $U_C$. $\omega_D$ and $\omega_c$ are the characteristic energy cutoff of the

Debye phonon and non-phonon respectively. Within our assumption, the interaction potential $V_{kk'}$ may be written as

$$V_{kk'}(\omega) = (-V_{ph} - U_C)f(k)f(k') \quad , \quad 0 < |\varepsilon_k, \varepsilon_{k'}| < \omega_D$$
$$= (-U_C)f(k)f(k') \quad , \quad \omega_D < |\varepsilon_k, \varepsilon_{k'}| < \omega_c$$

(2)

For such as the interaction, the superconducting order parameters are

$$\Delta_{jk} = \Delta_{j1}(T)f(k) \quad \text{for} \quad 0 < |\varepsilon_k| < \omega_D$$
$$= \Delta_{j2}(T)f(k) \quad \text{for} \quad \omega_D < |\varepsilon_k| < \omega_c$$

(3)

that $j = 1, 2$. Here $\Delta_{jk}$ represents the temperature and wave vector dependent order parameter., $\Delta_{j1}$, $\Delta_{j2}$ are the phonon and non-phonon parts of the order parameters that are not dependent on wave vector.

We can get the order parameter equation in the vicinity of critical temperature ($T_c$) with constant density of state ($N(0)$) by substitution Eqs.(2) and (3) into Eq.(1)

$$\begin{bmatrix} \Delta_{11} \\ \Delta_{21} \\ \Delta_{12} \\ \Delta_{22} \end{bmatrix} = \begin{bmatrix} (\lambda_1 + \mu_1)I_1(\Delta_1) & (\lambda_{12} + \mu_{12})I_1(\Delta_2) & \mu_1 I_2 & \mu_{12} I_2 \\ (\lambda_{12} + \mu_{12})I_1(\Delta_1) & (\lambda_2 + \mu_2)I_1(\Delta_2) & \mu_{12} I_2 & \mu_2 I_2 \\ \mu_1 I_1(\Delta_1) & \mu_{12} I_1(\Delta_2) & \mu_1 I_2 & \mu_{12} I_2 \\ \mu_{12} I_1(\Delta_1) & \mu_2 I_1(\Delta_2) & \mu_{12} I_2 & \mu_2 I_2 \end{bmatrix} \begin{bmatrix} \Delta_{11} \\ \Delta_{21} \\ \Delta_{12} \\ \Delta_{22} \end{bmatrix}$$

(4)

Here, the anisotropic function of order parameter are changed into spherical coordinate that $f(k) \equiv f(\theta, \phi)$. We can get

$$I_1(\Delta_i) = \frac{1}{4\pi} \int_0^{2\pi} \int_0^{\pi} \sin\theta \, d\theta d\phi \int_0^{\omega_D} \frac{f^2(\theta,\phi)}{\sqrt{\varepsilon_k^2 + \Delta_i^2(T)f^2(\theta,\phi)}} \tanh\left(\frac{\sqrt{\varepsilon_k^2 + \Delta_i^2(T)f^2(\theta,\phi)}}{2T}\right) d\varepsilon_k$$

$$\cong <f^2(\theta,\phi)> \ln(\frac{2\omega_D \gamma}{\pi T}) - <f^4(\theta,\phi)> \Delta_i^2 \frac{7}{8} \frac{\xi(3)}{\pi^2 T^2}$$

(5.1)

$$I_2(\Delta_i) = \frac{1}{4\pi} \int_0^{2\pi} \int_0^{\pi} \sin\theta d\theta d\phi \int_{\omega_D}^{\omega_C} \frac{f^2(\theta,\phi)}{\sqrt{\varepsilon^2 + \Delta_i^2(T)f^2(\theta,\phi)}} \tanh\left(\frac{\sqrt{\varepsilon^2 + \Delta_i^2(T)f^2(\theta,\phi)}}{2T}\right) d\varepsilon_k$$

$$\cong <f^2(\theta,\phi)> \ln(\frac{\omega_C}{\omega_D})$$

(5.2)

that

$$<f^2(\theta,\phi)> = \frac{1}{4\pi} \int_0^{2\pi} \int_0^{\pi} f^2(\theta,\phi) \sin\theta \, d\theta d\phi$$

(5.3)

The coupling constants are defined as

$$\lambda_1 = N(0)V_{ph}^1, \lambda_2 = N(0)V_{ph}^2, \lambda_{12} = N(0)V_{ph}^{12} = N(0)V_{ph}^{21}$$

and
$$\mu_1 = N(0)U_C^1, \mu_2 = N(0)U_C^2, \mu_{12} = N(0)U_C^{12} = N(0)U_C^{21}$$

with $\lambda_{12} = \lambda_{21}$, $\mu_{12} = \mu_{21}$.

where $a_\lambda = \dfrac{\lambda_{12}^2 - \lambda_1\lambda_2}{\lambda_1 + \lambda_2} = \dfrac{\lambda_{12}^2 - \lambda_1\lambda_2}{\lambda_t}$ and $b_\mu = \dfrac{\mu_{12}^2 - \mu_1\mu_2}{\mu_1 + \mu_2} = \dfrac{\mu_{12}^2 - \mu_1\mu_2}{\mu_t}$.

Solving Eq.(4) for $I_1$, we can get

$$aI_1(\Delta_1)I_1(\Delta_2) + bI_1(\Delta_2) + cI_1(\Delta_1) + d = 0 \tag{6}$$

where

$$a = (\lambda_1\mu_2 + \lambda_2\mu_1 - 2\lambda_{12}\mu_{12} - \lambda_t a_\lambda - \mu_t b_\mu + I_2\lambda_t\mu_t b_\mu + I_2^2\lambda_t\mu_t a_\lambda b_\mu + I_2\lambda_t\mu_t a_\lambda)$$
$$b = (-\lambda_2 - \mu_2 + I_2\lambda_2\mu_t - I_2\mu_t b_\mu + I_2^2\lambda_2\mu_t b_\mu)$$
$$c = (-\lambda_1 - \mu_1 + I_2\lambda_1\mu_t - I_2\mu_t b_\mu + I_2^2\lambda_1\mu_t b_\mu)$$
$$d = 1 - I_2\mu_t - I_2^2\mu_t b_\mu$$

and $b + c = -m = \lambda_t + \mu_t - I_2\lambda_t\mu_t + 2I_2\mu_t b_\mu - I_2^2\lambda_t\mu_t b_\mu$

After some calculations, we can get the order parameter as

$$\Delta_1^2(T) = \frac{8\pi^2 T^2}{7\zeta(3)\langle f^4(\theta,\phi)\rangle} \frac{\left[m\langle f^2(\theta,\phi)\rangle\left(1 - \dfrac{T}{T_c}\right) - a\langle f^2(\theta,\phi)\rangle^2 2\left(1 - \dfrac{T}{T_c}\right)h(T_c)\right]}{\left[c + \alpha^2 b + a\langle f^2(\theta,\phi)\rangle(h(T_c) + \left(1 - \dfrac{T}{T_c}\right)(\alpha^2 + 1)\right]} \tag{7}$$

Here $h(T) = \ln(\dfrac{2\omega_D \gamma}{\pi T})$ and we define the ration of 2nd order parameter o to 1st order parameter to be a temperature independent parameter as $\alpha = \dfrac{\Delta_2(T)}{\Delta_1(T)}$.

Within our model, we can write the specific heat as

$$C(T) = \frac{2N_0}{T^2} \frac{1}{4\pi} \int_0^{2\pi}\int_0^\pi \sin\theta d\theta d\phi \int_{-\omega_D}^{\omega_D} \frac{\exp\left(\sqrt{\varepsilon_k^2 + (\Delta_1^2(T) + \Delta_2^2(T))f^2(\theta,\phi)}/T\right)}{\left[\exp\left(\sqrt{\varepsilon_k^2 + (\Delta_1^2(T) + \Delta_2^2(T))f^2(\theta,\phi)}/T\right) + 1\right]^2} \times$$
$$\left[\varepsilon_k^2 + (\Delta_1^2(T) + \Delta_2^2(T))f^2(\theta,\phi) - \frac{2}{T}f^2(\theta,\phi)\frac{d}{dT}(\Delta_1^2(T) + \Delta_2^2(T))\right]d\varepsilon_k \tag{8}$$

After solving with the conventional process, the specific heat jump equation is

$$\left.\frac{\Delta C}{C_N}\right|_{T=T_c} = 1.43 \frac{<f^2(\theta,\phi)>^2}{<f^4(\theta,\phi)>}(1+\alpha^2)\frac{(c + b + 2a <f^2(\theta,\phi)> h(T_c))}{(c + \alpha^2 b + a(\alpha^2+1)<f^2(\theta,\phi)> h(T_c))} \tag{9}$$

Eq.(9) is the specific heat jump equation of anisotropic two-band superconductors with the effect of phonon and non-phonon coupling constant and the anisotropic function of order parameter that can be reduced to anisotropic one-band superconductor[9,17,20-22] as $\left.\frac{\Delta C}{C_N}\right|_{T=T_c} = 1.43 \frac{<f^2(\theta,\phi)>^2}{<f^4(\theta,\phi)>}$ and BCS 's results as $\left.\frac{\Delta C}{C_N}\right|_{T=T_c} = 1.43$ .

The numerical calculation of Eq.(8) can be shown by assumed the anisotropic gap functions. There are two model consider the effect of anisotropic in c-axis with respect to ab-plane that agree with the physical properties of our consideration superconductors. They are Haas and Maki [15]'s model as $\Delta(k) = \Delta(0)\frac{1+a'z^2}{1+a'}$ and Posazhennikova, Dahm and Maki [16]'s model as $\Delta(k) = \frac{\Delta(0)}{\sqrt{1+a'z^2}}$ where $a'$ is the anisotropic parameter and $z = \cos\theta$, $\theta$ is the polar angle.

### 3. Results and discussions

The specific heat of $MgB_2$, $Lu_2Fe_3Si_5$, $Nb_3Sn$ superconductors are shown in Figure.(1-3) that the experimental data of Ref.[5-7] are compared to our numerical calculation. The anisotropic function of the Haas and Maki(HM)[15] and Posazhennikova, Dahm and Maki (PDM) [16] are use for our calculations. There are two kinds of the parameters used in our calculations, the experimental results and the modeling assumption parameters.

For $MgB_2$ superconductor, the experimental data are chosen from that presented in Ref.[5] as $\omega_D = 750\,K$, $T_c = 38\,K$, and $\alpha = 0.33$ ($\alpha = \frac{\Delta_2(T)}{\Delta_1(T)} \approx \frac{\Delta_2(0)}{\Delta_1(0)} = \frac{1.3}{3.9}$). The modeling assumption parameters are $\omega_C = 800\,K$, $\lambda_2 = 0.1$, $\mu_1 = 0.5$, $\mu_2 = 0.1$, $\mu_{12} = 0.1$, $\lambda_{12} = 0.1$ for both model except the $\lambda_1$ that $\lambda_1 = 0.58$ for HM and $\lambda_1 = 0.51$ for PDM. In Figure.1, our results show $\left.\frac{\Delta C}{C_N}\right|_{T=T_c}$ equal to 0.89 and 0.86 for HM and PDM that agreed with experimental value 0.82. For $Lu_2Fe_3Si_5$ superconductor, the experimental data presented in Ref.[6] are chosen that $T_c = 6.1\,K$, and $\alpha = 0.25$ ($\alpha = \frac{\Delta_2(T)}{\Delta_1(T)} \approx \frac{\Delta_2(0)}{\Delta_1(0)} = \frac{1.1}{4.4}$) and the modeling assumption parameters are $\omega_D = 450K$, $\omega_C = 750K$, $\lambda_2 = 0.2$, $\mu_1 = 0.2$, $\mu_2 = 0.2$, $\mu_{12} = 0.1$, $\lambda_{12} = 0.1$ for both model except the $\lambda_1$ that $\lambda_1 = 0.47$ for HM and $\lambda_1 = 0.41$ for PDM. In Figure 2, the $\left.\frac{\Delta C}{C_N}\right|_{T=T_c}$ are equal to 1.04, 1.06, and 1.05 for HM, PDM and experiment data. For $Nb_3Sn$ superconductor, we choose the experimental data from Ref.[7] that $T_c = 17.8K$, $\omega_D = 230K$ and $\alpha = 0.16$ ($\alpha = \frac{\Delta_2(T)}{\Delta_1(T)} \approx \frac{\Delta_2(0)}{\Delta_1(0)} \approx \frac{0.8}{4.9}$) .The modeling assumption parameters are $\omega_C = 330\,\text{K}$ , $\lambda_2 = 0.5, \mu_1 = 0.5, \mu_2 = 0.5, \mu_{12} = 0.2$ , $\lambda_{12} = 0.2$ for both model except the $\lambda_1$ that $\lambda_1 = 0.47$ for HM and $\lambda_1 = 0.41$ for PDM. The $\left.\frac{\Delta C}{C_N}\right|_{T=T_c}$ are equal to 2.59 , 2.64, and 2.55 for HM, PDM, and experiment that shown in Figure.3.

We find that our calculations can describe the specific heat of $MgB_2$, $Lu_2Fe_3Si_5$, $Nb_3Sn$ superconductors for $T \approx T_c$ well. The experimental result of $Nb_3Sn$ superconductor is best fit to our model for all range of temperature. Within our parameters, we find that $Nb_3Sn$ is almost the isotropic superconductors and there is the effect of interband interaction on the specific heat jump. For two-band model, the highest value of specific heat jump for isolated band is ; $1.43 + 1.43 = 2.86$ ; and $Nb_3Sn$ has 2.55 . And the worst fit to our model is $Lu_2Fe_3Si_5$ superconductor that can fit only at the specific heat jump because this superconductor has the magnetic order in superconducting state and we do not include in our calculation.

**4.Conclusions**

The effect of anisotropic order parameters to specific heat of two band superconductors in BCS weak-coupling limit are investigated. The experimental data of $MgB_2$ , $Lu_2Fe_3Si_5$ and $Nb_3Sn$ are compared to our numerical calculations with the anisotropic function of HM and PDM .We find that for near critical temperature, our results can describe experimental data of two-band superconductors. The $Nb_3Sn$ superconductor show almost isotropic superconductors properties that best fit to our model. The $MgB_2$ superconductor can fit only near critical temperature well and $Lu_2Fe_3Si_5$ superconductor can fit only at critical temperature .Two models of anisotropic order parameters are used for numerical calculations that we find little effect to the numerical results when compared to the experimental data. The difference between two models is the modeling assumption parameters.

**Acknowledgement** :The author would like to thank Professor Dr.Suthat Yoksan for the useful discussion and also thank Srinakharinwirot University, the university of the Thai Chamber of Commerce, and ThEP Center for the financial support.

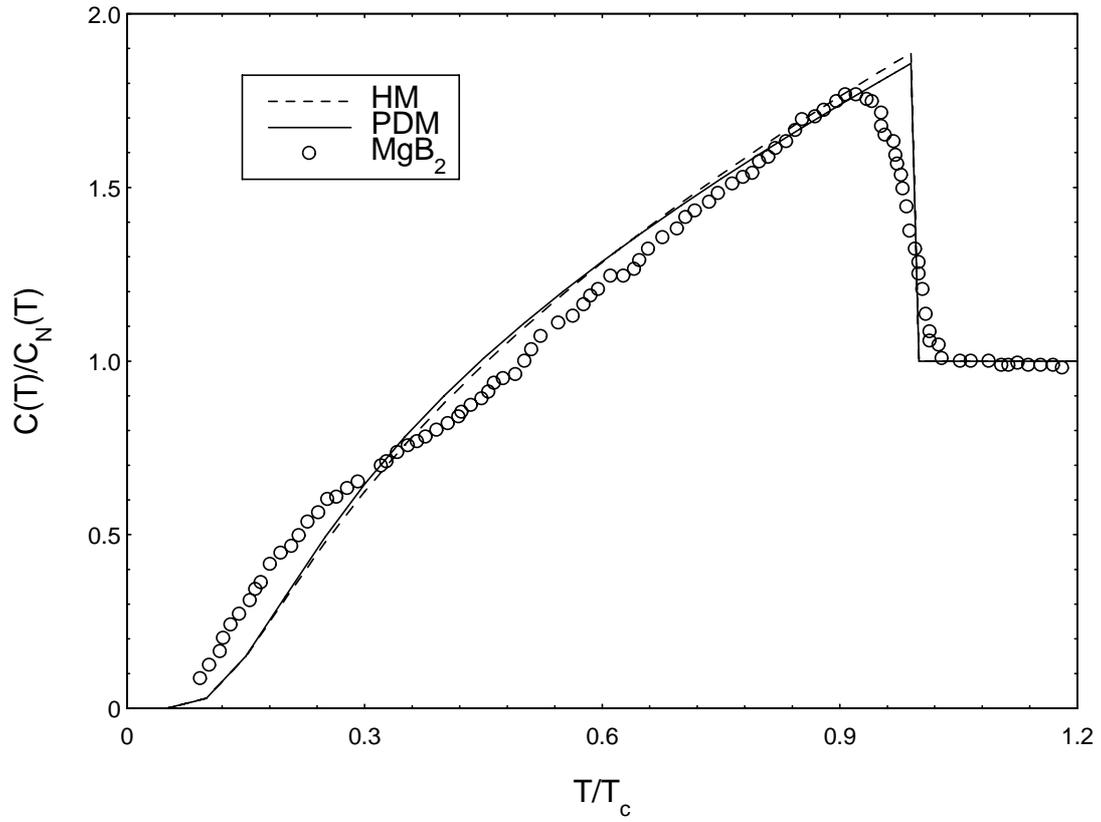

**Figure(1).** Our calculations (dash and solid line) and experimental data[5]("o") of the $MgB_2$ superconductor versus temperature is shown. Here we use the anisotropic function of HM [15] and PDM [16].

9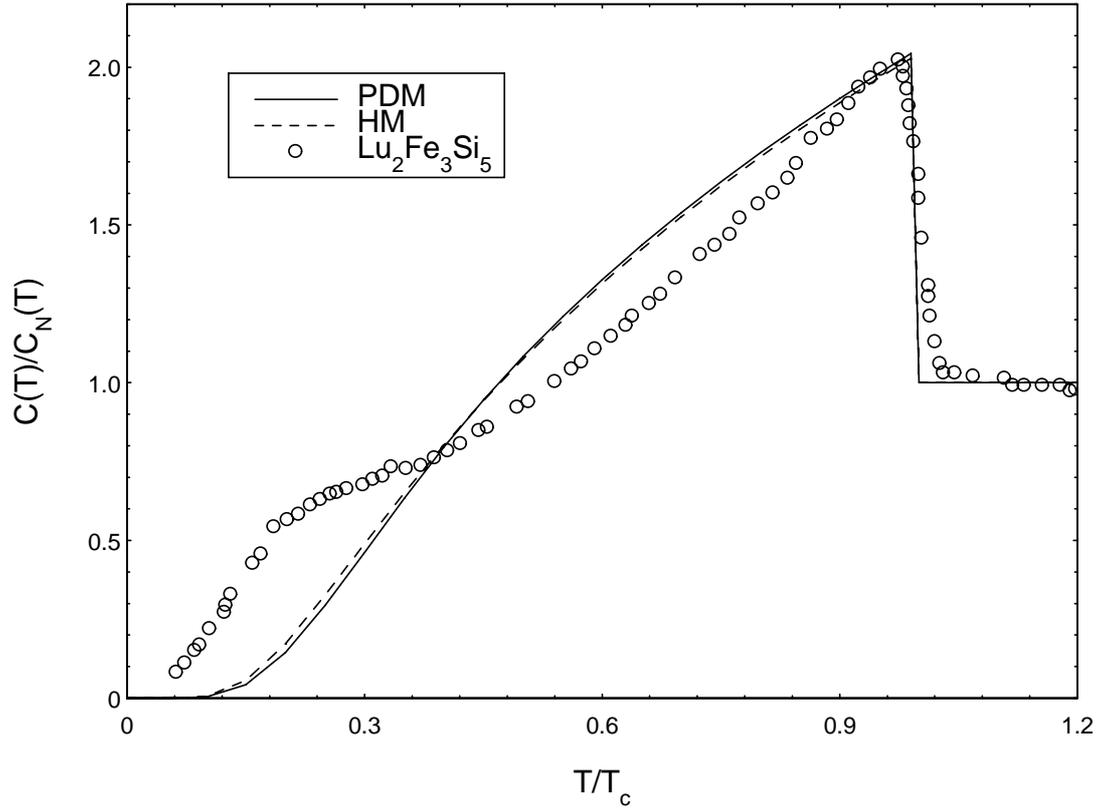

**Figure.(2).** Our calculations (dash and solid line) and experimental data[6]("o") of the $Lu_2Fe_3Si_5$ superconductor versus temperature is shown . .Here we use the anisotropic function of HM [15] and PDM [16].

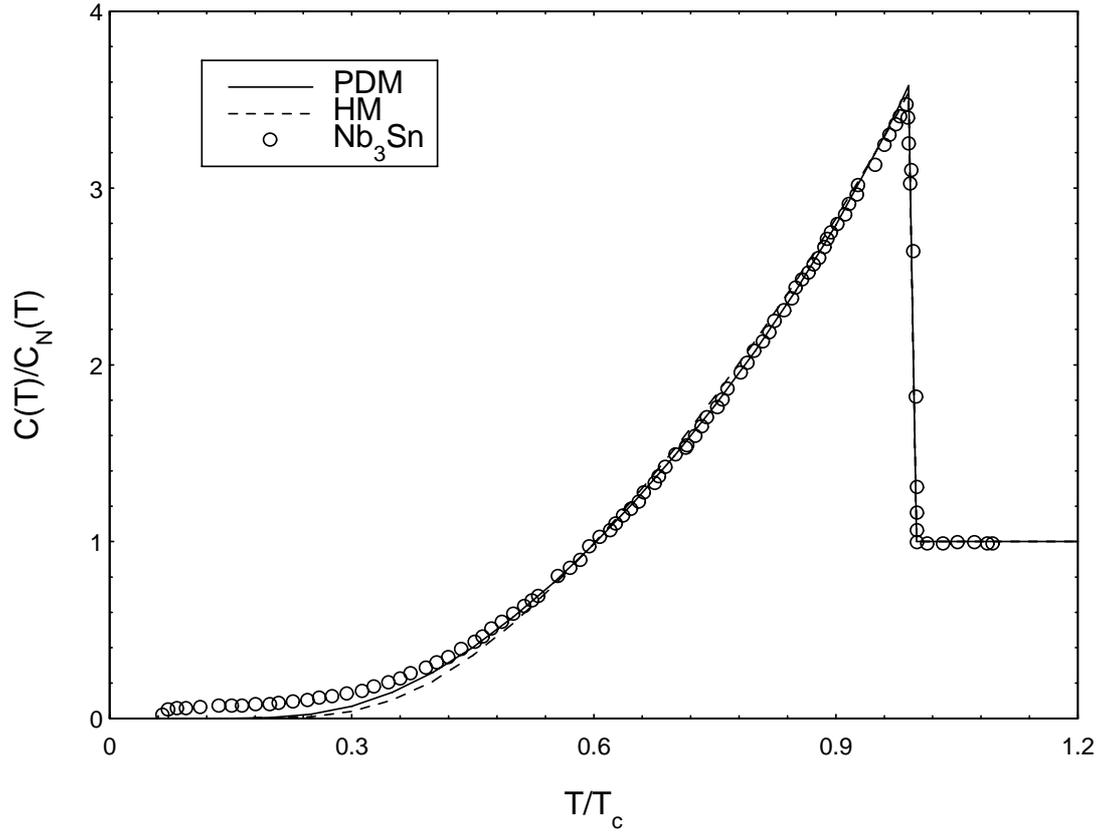

**Figure.(3).** Our calculations (dash and solid line) and experimental data[7]("o") of the $Nb_3Sn$ superconductor versus temperature is shown .Here we use the anisotropic function of HM [15] and PDM [16].